\documentclass[preprint,preprintnumbers,amsmath,amssymb,superscriptaddress]{revtex4-1}
\usepackage{fullpage,amsthm,amsmath,amsfonts,bm,times,color,bbm}
\usepackage{graphicx,subfigure}
\usepackage[english]{babel}
\usepackage{mathtools}
\usepackage{hyperref}

\def\eqa{\begin{eqnarray}}
\def\eea{\end{eqnarray}}
\newcommand{\eq}{\begin{equation}}
\newcommand{\ee}{\end{equation}}

\renewcommand{\>}{\rangle}

\makeatother

\makeatletter

\begin{document}

\title{A combined network and machine learning approaches for product market forecasting
 }

\author{Jingfang Fan}
\email{j.fang.fan@gmail.com}
\affiliation{Department of Physics, Bar Ilan University, Ramat Gan 52900, Israel}
\affiliation{Potsdam Institute for Climate Impact Research, Potsdam 14412, Germany}
\author{Keren Cohen}
\affiliation{Department of Mathematics, Bar Ilan University, Ramat Gan 52900, Israel}
\author{Louis M. Shekhtman}
\affiliation{Department of Physics, Bar Ilan University, Ramat Gan 52900, Israel}

\author{Sibo Liu}
%\email{siboliu@ln.edu.hk}
\affiliation{Department of Economics, Lingnan University, Tuen Mun, Hong Kong}

\author{Jun Meng}
%\email{j.fang.fan@gmail.com}
\affiliation{Department of Physics, Bar Ilan University, Ramat Gan 52900, Israel}
\affiliation{Potsdam Institute for Climate Impact Research, Potsdam 14412, Germany}
\author{Yoram Louzoun}
\email{louzouy@math.biu.ac.il}
\affiliation{Department of Mathematics, Bar Ilan University, Ramat Gan 52900, Israel}
\author{Shlomo Havlin}
%\email{havlin@ophir.ph.biu.ac.il}
\affiliation{Department of Physics, Bar Ilan University, Ramat Gan 52900, Israel}

\begin{abstract}
Sustainable financial markets play an important role in the functioning of human society. Still, the
detection and prediction of risk in financial markets remain challenging and draw much attention from the scientific community. Here we develop a new approach based on combined network theory and machine learning to study the structure and operations of financial product markets. Our network links are based on the similarity of firms' products and are constructed using the Securities Exchange Commission (SEC) filings of US listed firms. We find that several features in our network can serve as good precursors of financial market risks. We then combine the network topology and machine learning methods to predict both successful and failed firms. We find that the forecasts made using our method are much better than other well-known regression techniques. The framework presented here not only facilitates the prediction of financial markets but also provides insight and demonstrate the power of combining network theory and machine learning.

\end{abstract}
\date{\today}

%\flushbottom
\maketitle
\section{INTRODUCTION}

Network science has been used to predict many natural and technological phenomena. Examples include, the evolution of a scientist's impact \cite{sinatra_quantifying_2016}, forecast disease epidemics \cite{eubank_modelling_2004,colizza2006role,brockmann_hidden_2013}, predict the spatial development of urban areas~\cite{li_simple_2017} and 
forecast climate extreme events \cite{ludescher_very_2014,boers_prediction_2014,meng_percolation_2017,meng_forecasting_2018}.
Network approach demonstrated its potential as a useful tool in the study of real world systems, such as, physics, biology, and social systems \cite{newman2010networks,cohen2010complex,zhao_inducing_2013}. Recently,  a network approach has been applied to describe the instability in financial systems~\cite{bardoscia_pathways_2017}, and to study the relationship between the structure of the financial network and the likelihood of systemic failures due to contagion of risk~\cite{miura_effect_2012,acemoglu_systemic_2015}.

%Machine learning as an artificial-intelligence method for analysing data is becoming more
%and more prevalent in an increasing number of fields \cite{butler_machine_2018}. This is due
%to a combination of the availability of big data and
%advances in computational power. 
%Two typical methods of machine learning can be distinguished, namely the unsupervised and supervised methods \cite{lecun_deep_2015}.

In the present study, we use text-based analysis of SEC Form 10-K product descriptions to 
construct the network of product similarity between firms. The firms are regarded as network nodes, and the level of similarity between
the product descriptions of different firms represents the network links (strength). The 10-K product descriptions are obtained from
the Securities Exchange Commission (SEC) Edgar website (\url{https://www.sec.gov/}). We then analyze the topological structure of the network, and determined how measures such as, clustering coefficient
correlate with membership in the Standard \& Poor's 500 (S\&P 500). Furthermore, we also analysed the $K$-shell structure of the network \cite{carmi_model_2007,kitsak_identification_2010} and find that it reveals that firms in more outer shells have higher risk to collapse (or merge). Furthermore, we combine the network structure and machine learning methods to predict both the successful and collapsed firms. We find that the forecasting rates by using our combined method are significantly higher than random guessing and other methods \cite{breiman_random_2001}.

\section{Data}
\subsection{Product Similarity Data}
In this study, we use text-based analysis of 10-K product descriptions to obtain the product similarity between firms, representing the links.
For any two firms $i$ and $j$, we have a product similarity, which is a real value in the
interval [0,1] describing the similarity of words used by firms $i$ and $j$ in their 10-K forms.
To compute the  ``product similarity'' between two firms using the basic cosine similarity method ~\cite{hoberg_product_2010,Hoberg2016jpe},
We first build the database for each year by taking the list of unique words used in all product descriptions in that year.
We then take the text from each firm's product description and construct a binary $N$-vector summarizing its word usage. A given element of this
$N$-vector is 1 if the given dictionary word is used in firm $i$'s product description. For each firm $i$, we denote this binary
$N$-vector as $\vec{P_{i}}$. We define the normalized vector $\vec{V_i}$ as,
\begin{equation}
\vec{V_{i} }= \frac{\vec{{P_{i}}}}{\sqrt{\vec{P_{i}}\cdot\vec{ P_{i}}}}.
\label{eqa1}
\end{equation}

To measure the similarity of products of firms $i$ and $j$, we compute the dot product of their normalized vectors, which is then the basic cosine similarity:

\begin{equation}
w_{i,j} = \frac{\vec{V_{i}}\cdot\vec{V_{j}}}{\mid V_{i} \mid \mid V_{j} \mid}.
\label{eqa2}
\end{equation}
In this study, we use 18 years (from 1996 to 2013) of data. For a more detailed description see Ref. \cite{hoberg_product_2010}.

\subsection{External Financial Data}
Apart from the product similarity data, we also compile some external information to construct financial variables. We download the daily closing price of the S\&P 500 index from the Center for Research in Security Prices (CRSP) (\url{ http://www.crsp.com/}). This time-series  can be used to measure the overall market dynamics. Several firm-specific variables are also taken to control the range of observed firm traits. Specifically, firm size is assumed to be the total market capitalization following as suggested in Ref. \cite{fama1992cross}; the book-to-market ratio is the book value of equity over the market value of equity and measures a firm's growth opportunities; leverage is a measure of capital structure defined by the ratio of long-term debt and other debt (current liabilities) to the sum of long-term debt, debt in current liabilities, and stockholders' equity; profitability is defined as income before extraordinary items over lagged total assets; prior year return is the return of the stock in the past year; investment is the year-over-year percent growth in total assets following \cite{cooper2008asset}; liquidity is a measure of firm liquidity found using reference \cite{amihud2002illiquidity}; and the Altman Z-score is a measure of default risk according to reference \cite{altman1968financial}. The data can be downloaded from the Compustat database (\url{http://www.compustat.com}).

\section{Methods}
\subsection{Product Similarity Network}
\begin{figure}
\begin{centering}
\includegraphics[width=1\textwidth]{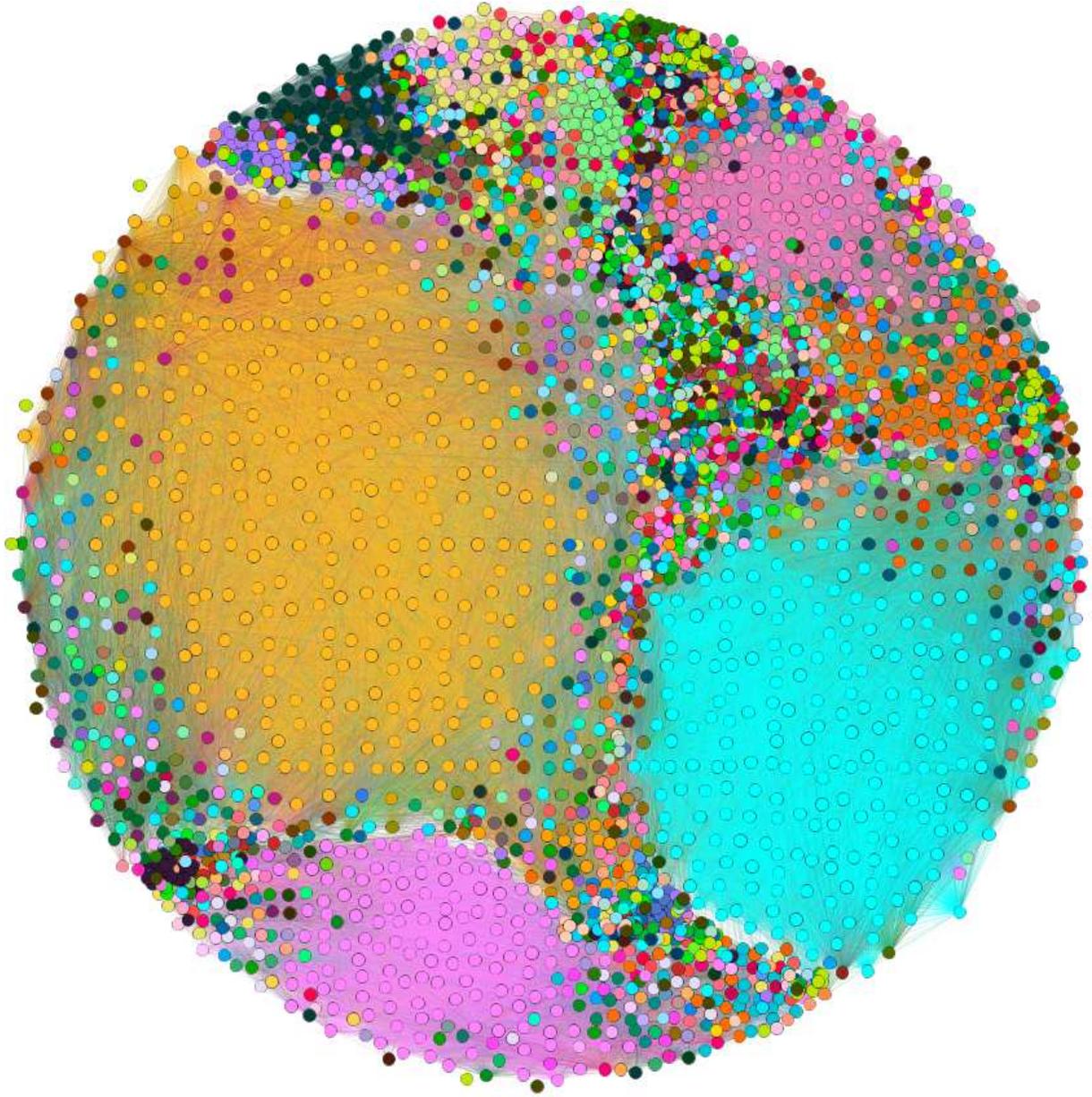}
\caption{\label{fig1} A typical snapshot of the product similarity network for the year 2012.  The system size is $N = 3925$, average degree is $\<k\> = 94.78$. Different colors represent different shells.}
\par\end{centering}
\end{figure}

For each year, we construct a weighted undirected network based on the product similarity data where each firm or company is a node and the links have a given strength, $w$, representing the level of similarity between pairs of nodes.  The links are thresholding such that only links with a weight greater than significant value are kept in the network. Here, we only consider the similarity values that are above $10^{-4}$.  This threshold is calculated  based on the coarseness of the three-Digit Standard Industrial Classification (SIC). The level of coarseness thus matches 
that of three digit SIC codes, as both classifications result in the same number of firm pairs being deemed related \cite{hoberg_product_2010}.  We present a specific product similarity network for the year 2012 in Fig.~\ref{fig1}.

\subsection{Network topology measures for machine learning}
Multiple network topological measures have been used to predict the future gain of companies, as well as the probability of their collapse. The general approach is based on methods described in Rosen \textit{et. al} \cite{rosen_topological_2016} and Naaman \textit{et. al} \cite{naaman_edge_nodate}. In brief, a vector representing a set of topological features has been computed for each node, and this vector was then used in a machine learning framework as decribed below. The following features were used for the topological network attribute vector (NAV):
\begin{itemize}
\item Degree (number of neighbors).
\item Betweenness Centrality \cite{everett_centrality_1999}. Betweenness is a centrality measure of a node (or a link). It is defined as the number of shortest paths between all vertex pairs that pass through the node.
\item Closeness Centrality \cite{sabidussi_centrality_1966}. Closeness is a centrality measure of a node (or a link) defined as the average length of the shortest path between the vertex and all other vertices in the graph.
\item Distance distribution moments. We compute the distribution of distances from each node to all other nodes using a Dijkstra's algorithm \cite{dijkstra_note_1959}, and then take the first and second moments of this distribution. 
\item Flow \cite{rosen_directionality_2014}. We define the flow a node as the ratio between the directed and undirected distances between this node and all other nodes. 
\item Network motifs \cite{milo_network_2002}. Network motifs are patterns of small connected sub-graphs. We use an extension of the Itzchack  algorithm \cite{itzhack_optimal_2007} to calculate motifs. For each node, we compute the frequency of each motif in which this node participates. Note that the above algorithm is an extension of the concept to undirected graphs. 
\item  $K$-core \cite{batagelj_fast_2011}.  The $K$-core of a network is a maximal subgraph that contains only vertices with degree $k$ or more. Equivalently, it is the subgraph of $G$ formed by recursively deleting all nodes of degree less than $k$.

\end{itemize}
\subsection{Machine learning}
We computed for each node an NAV as described above. All nodes within a given year were then split into a training set composed of $70$\% of nodes and a test set composed of the remaining $30$\%. Each node was also associated with two binary valued tags representing whether the company collapsed in the following year and if it was within the top $5$\% of returns in the following year. The values of the tags were then learned using either a Random Forest (RF) \cite{ho_random_1995} classifier and a neural network with two internal layers. The Random Forest was trained with $200$ trees and a balanced loss matrix, where the error cost was inversely proportional to the training fraction of the appropriate tag (0 or 1). The trees were limited to $10$ samples per leave. All other parameters were the default parameters of the Matlab Treebagger. 
The neural network had a rectified linear unit (ReLU) activation functions and a cross entropy loss function.
%The size of the internal layers were -------------- and --------------. 
The solver used was an ADAM optimizer \cite{kingma_adam:_2014}. L2 regularization has been also applied.  All other parameters were as in the default of the Keras python library: \url{https://keras.io}.

\section{Results}

\begin{figure}
\begin{centering}
\includegraphics[width=1\textwidth]{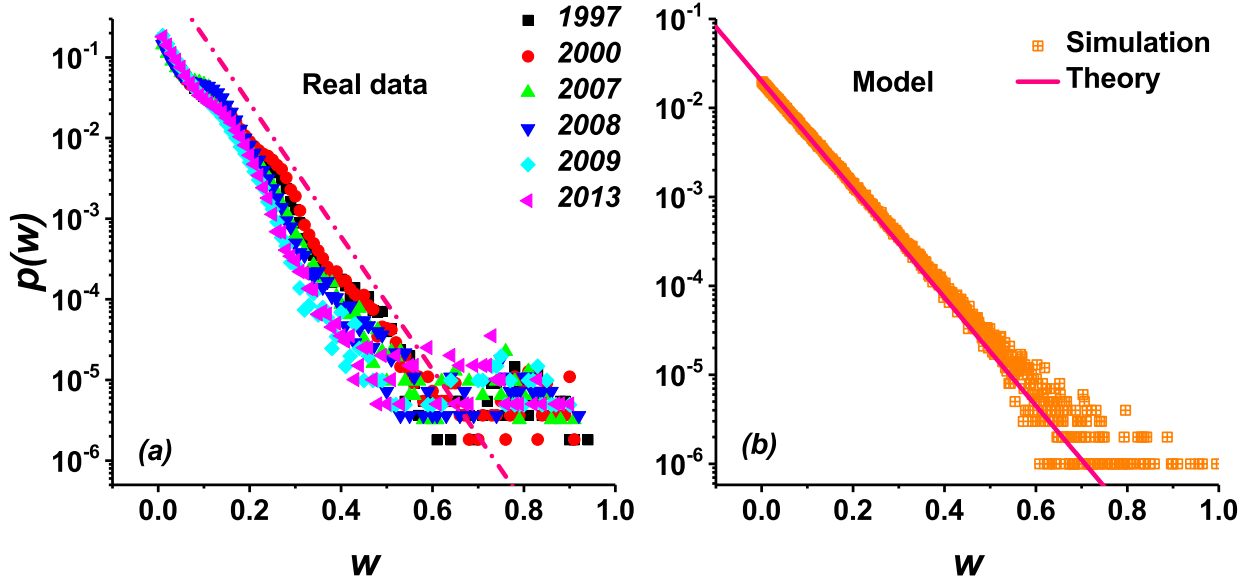}
\caption{\label{fig2} The probability distribution functions of links weights (a) for real networks of different years, the dashed line is a fitting line using an exponential distribution; (b) Simulation and theoretical results for our competitive-cooperative model.}
\par\end{centering}
\end{figure}

We first consider the probability density function (PDF) of links' strength, $p(w)$, for each network (year). The results for six specific years are shown in Fig.~\ref{fig2} (a). We find that $p(w)$ is robust and approximately follows an exponential distribution. The values of product similarity can reflect the product market synergies and competition of mergers and acquisitions \cite{hoberg_product_2010}, i.e., higher $w$ may mean the two firms are highly competitive or in cooperative relations. To better understand and explain the exponential distribution, we develop a competitive-cooperative model. For each node (firm) $i$, we set a score $x_{i}$, which follows a general mathematical function $f(x)$. %{\color{red} Unclear what this score is or where it comes from}. 
We define the weight/strength relation between nodes $i$ and $j$ as,
\begin{equation}
w_{i,j} = -\Lambda \log[1 - |x_{i} -x_{j}|],
\label{eqa3}
\end{equation}
where $|x_{i} -x_{j}|$ is the absolute difference between $i$ and $j$, i.e., high values indicating a large difference or low similarity. 
$\Lambda$ is a constant used to guarantee that $w_{i,j}$ is between $[0,1]$. If we assume that $f(x)$ follows an uniform random distribution, we can show in our model that $p(w)$ follows an exponential distribution (see more details in SI \cite{SI}).
Indeed,  Fig.~\ref{fig2}(b) shows our Monte Carlo simulations and theoretical results of the model, where we find that $p(w)$ agrees well with the real data. It suggests that our competitive-cooperative model can be used to accurately simulate/describe the real product similarity network. 

%{\color{red} This section is very confusing and it is unclear what $f(x)$ represents. It is also not clear to me if it follows our main message.}

Next, we consider two basic parameters to reveal the structure of the network: the weighted degree and clustering coefficient \cite{watts_collective_1998}. The weighted degree for node $i$ is defined as,
\begin{eqnarray}
s_{i} = \sum\limits_{j=1}^{N} a_{ij} w_{ij},
\label{eqa4}
\end{eqnarray}
where $a_{ij}$ is the adjacency matrix and $s_i$ quantifies the strength of node $i$ in terms of the total weight of its connections. In the case of the product similarity networks, it reflects the importance or impact of a firm $i$ in the network. We find, see Fig.~\ref{Fig:S1}, that the strength $s(k)$ of nodes with degree $k$ increases with $k$ as,
\begin{eqnarray}
%%\label{11}
s \sim k^{\beta}.
\label{eqa5}
\end{eqnarray}
We find that the
power-law fit for the real data gives an exponent $\beta \approx 1.5$. This value implies that the strength of nodes grows faster than their degree, i.e., the weight of edges belonging to highly connected nodes tends to have a higher value.
We notice that the universal power--law relationship is also observed in other real networks, e.g., the world-wide airport network, even with the same value of $\beta$ \cite{barrat_architecture_2004}.

\begin{figure}
\begin{centering}
\includegraphics[width=1\textwidth]{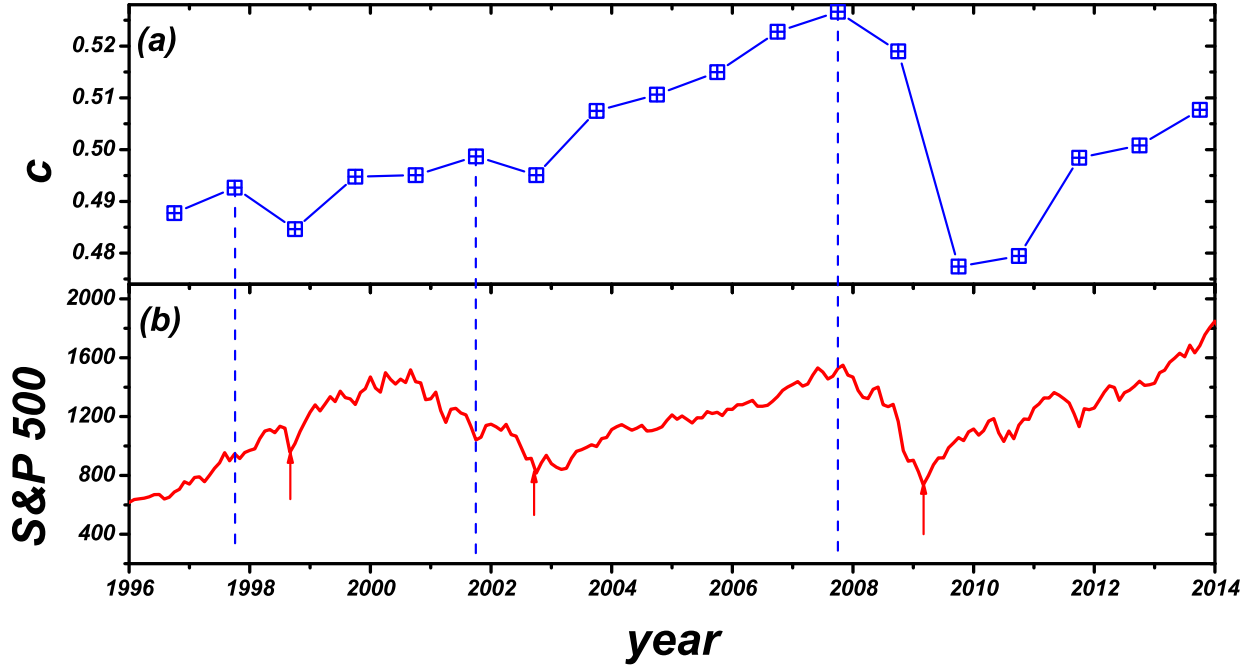}
\caption{\label{fig3} (a) Evolution of the clustering coefficient $C$ with years. (b) The evolution of the S\&P 500 Index.}
\par\end{centering}
\end{figure}

For weighted networks, the clustering coefficient for node $i$ is defined as the geometric average of the subgraph edge weights \cite{saramaki_generalizations_2007},
\begin{eqnarray}
c_{i}  = \frac{1}{k_{i}(k_{i}-1)} \sum\limits_{j,k} (\hat{w}_{i,j} \hat{w}_{i,k} \hat{w}_{j,k})^{1/3},
\label{eqa6}
\end{eqnarray}
where the edge weight is normalized by the maximum weight in the network, $\hat{w}_{i,j} = w_{i,j}/max(w)$. The average clustering coefficient is $
C  = \frac{1}{N} \sum\limits_{i=1}^{N} c_{i}$, representing the presence of triplets in the network.
Fig.~\ref{fig3} (a) shows the dynamical evolution of $C$ with time (years), 
Fig.~\ref{fig3} (b) shows the S\&P 500 Index. We find that the behavior of $C$ (1 year ahead) and the S\&P 500 Index is highly correlated (see Fig.~\ref{Figs:S2}). In particular, we can see 3 local maxima $C$ (labeled by dashed blue lines) and 3 local minima S\&P 500 (labeled by red lines with  arrows), which stands for three financial crisis: the 1997-1998 Asian Financial Crisis, the 2002 dotcom bubble  and the 2008 global financial crisis. The maxima values are always one year before the minima. This suggests that our network 
index $C$ might be able to help in forecasting the following year's stock market returns.

Given the relation between the evolved topology of the network of market, we tested whether the network contains enough information to predict at the single node level whether the company represented by this node will collapse or make exceptional gains in the following year.  We define the systemic risk as the disappeared firms (bankrupt or mergers and acquisitions) in the next year and the systemic return as the firms with the highest stock-market return ratio in the next year (we choose here the top $5\%$ of firms). Note that merger and bankrupt are obviously different. However,
most such events are bankruptcy and merger events. Thus, the events can be approximated as
bankruptcy.

To further show that network topology matters, we next analyze the $K$-shell structure of the network and find that the $K$-shell method is quite useful for predicting well above random, the disappeared firms. We present 
the receiver operating characteristic curve (ROC) using the $K$-shell method for predicting the disappeared firms of the 2012 network in Figs.~\ref{Figs:S3}. The area under the ROC curve (AUC) is significantly higher than the random case for all years (shown in Fig.~\ref{Figs:S4}). In particular, firms in the lower shells have higher market risk ratio. A possible reason is that the nodes in lower shells in the network are very fragile with higher risk. We present the scatter plot of risk ratio, $\rho$ (the ratio of disappeared firms at each shell), as a function of $K$-shell, see Fig.~\ref{Figs:S5} for three specific years, 1996, 2003 and 2012. Our results reveal that firms in more outer shells have higher risk to collapse.

\begin{figure}
\begin{centering}
\includegraphics[width=1\textwidth]{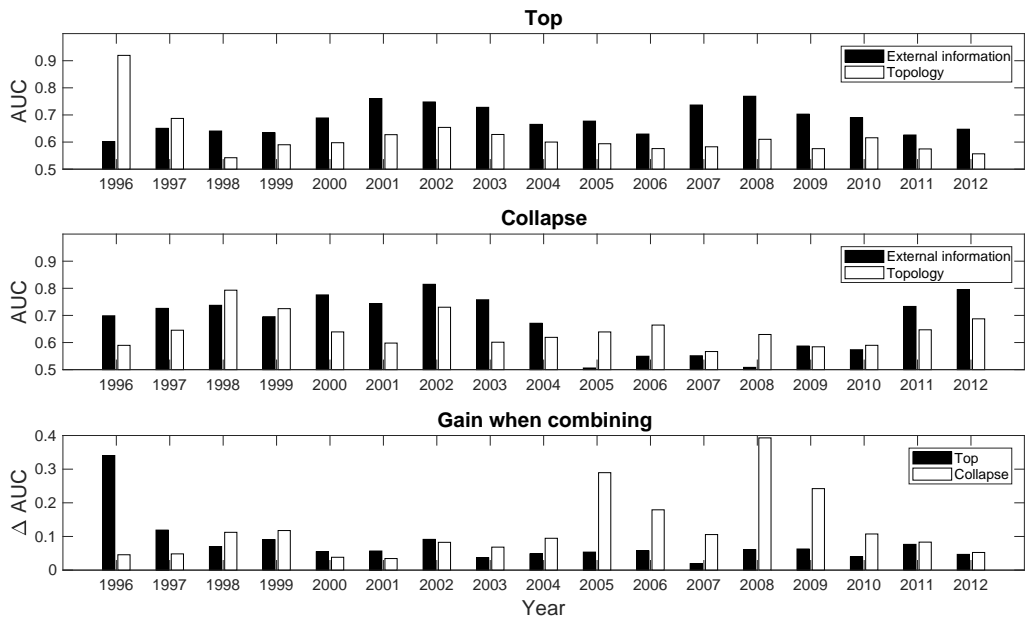}
\caption{\label{fig5} (Upper and Middle) Comparison of AUC for logistic regression on external features and Random Forest Classifier on topological features for both collapsed and top companies. (Bottom) The gain in AUC for Random Forest on combined topological and external features for the collapsed and top companies.}
\par\end{centering}
\end{figure}

Next a combined topological approach was tested to predict the same tags (collapsed or top) using a large set of topological measures and a Random Forest classifier. The combined features approach \cite{rosen_topological_2016} significantly  outperformed the $K$-Core based classification ( Fig.~\ref{fig5}). 
For both collapsed and top companies, we find that the topological information of our network provides significantly better predictions than the random case (the AUC is shown in Fig. \ref{fig5} upper and middle plots) for all 17 years. We also compared the AUC of machine learning using the network topology to the logistic regression based on the standard financial measures (non-network information): firm size, book-to-market ratio, leverage, profitability, prior year return, investment, liquidity, and Altman Z-score. The results are  shown in Fig. \ref{fig5} upper and middle plots.

The topological classifier is approximately of equal quality and often outperforms the logistic regression methods (Fig. \ref{fig5} upper and middle plots). Moreover when topological features are combined with the external features, the precision accuracy of RF method is significantly higher in almost all years and in both categories (see gain AUC in Fig. \ref{fig5} bottom plot). This highly suggests that the network topological information plays a prominent role in the prediction of future collapsed and top companies, it can be complementary to the traditional logistic regression methods. 
%
%To check that this was not a result of using different learning methods, we repeated the learning using Random Forest for both classical features and topology, with similar results. Thus, beyond its general interest the proposed network, the product feature network can actually have wide applications in analytic methods.

\section{SUMMARY}
In summary, we developed a combined network and machine learning algorithm to predict both collapsed and top companies in the financial market. Our network is based on the 10-K product similarities. We find that multiple topological measures inherent in our networks can serve as good precursors for machine learning approach (Random Forest). The forecasting rates using our method are often higher than using well-known logistic regression techniques. Moreover, when combining the external features and network topological measures, we find that the accuracy of the machine learning is significantly higher in almost all years and in both categories (collapsed and top gain).
The proposed method and analysis can provide
a new perspective for prediction of individual companies in the financial product markets and can potentially be used as a template to study other financial systems.

\section*{Acknowledgements}
We acknowledge the Italy-Israel project OPERA, the Israel-Italian collaborative project NECST, the Israel Science Foundation, ONR, Japan
Science Foundation, BSF-NSF, and DTRA (Grant no.
HDTRA-1-10-1-0014), EPICC for financial support.
PIK is a Member of the Leibniz Association.
\bibliography{MyLibrary}

\begin{thebibliography}{10}
\expandafter\ifx\csname url\endcsname\relax
  \def\url#1{\texttt{#1}}\fi
\expandafter\ifx\csname urlprefix\endcsname\relax\def\urlprefix{URL }\fi
\providecommand{\bibinfo}[2]{#2}
\providecommand{\eprint}[2][]{\url{#2}}

\bibitem{sinatra_quantifying_2016}
\bibinfo{author}{Sinatra, R.}, \bibinfo{author}{Wang, D.},
  \bibinfo{author}{Deville, P.}, \bibinfo{author}{Song, C.} \&
  \bibinfo{author}{Barab{\'a}si, A.-L.}
\newblock \bibinfo{title}{Quantifying the evolution of individual scientific
  impact}.
\newblock \emph{\bibinfo{journal}{Science}} \textbf{\bibinfo{volume}{354}},
  \bibinfo{pages}{5239} (\bibinfo{year}{2016}).
\newblock
  \urlprefix\url{http://science.sciencemag.org/content/354/6312/aaf5239}.

\bibitem{eubank_modelling_2004}
\bibinfo{author}{Eubank, S.} \emph{et~al.}
\newblock \bibinfo{title}{Modelling disease outbreaks in realistic urban social
  networks}.
\newblock \emph{\bibinfo{journal}{Nature}} \textbf{\bibinfo{volume}{429}},
  \bibinfo{pages}{180} (\bibinfo{year}{2004}).
\newblock \urlprefix\url{https://www.nature.com/articles/nature02541}.

\bibitem{colizza2006role}
\bibinfo{author}{Colizza, V.}, \bibinfo{author}{Barrat, A.},
  \bibinfo{author}{Barth{\'e}lemy, M.} \& \bibinfo{author}{Vespignani, A.}
\newblock \bibinfo{title}{The role of the airline transportation network in the
  prediction and predictability of global epidemics}.
\newblock \emph{\bibinfo{journal}{Proceedings of the National Academy of
  Sciences of the United States of America}} \textbf{\bibinfo{volume}{103}},
  \bibinfo{pages}{2015--2020} (\bibinfo{year}{2006}).

\bibitem{brockmann_hidden_2013}
\bibinfo{author}{Brockmann, D.} \& \bibinfo{author}{Helbing, D.}
\newblock \bibinfo{title}{The {Hidden} {Geometry} of {Complex},
  {Network}-{Driven} {Contagion} {Phenomena}}.
\newblock \emph{\bibinfo{journal}{Science}} \textbf{\bibinfo{volume}{342}},
  \bibinfo{pages}{1337--1342} (\bibinfo{year}{2013}).
\newblock \urlprefix\url{http://science.sciencemag.org/content/342/6164/1337}.

\bibitem{li_simple_2017}
\bibinfo{author}{Li, R.} \emph{et~al.}
\newblock \bibinfo{title}{Simple spatial scaling rules behind complex cities}.
\newblock \emph{\bibinfo{journal}{Nature Communications}}
  \textbf{\bibinfo{volume}{8}}, \bibinfo{pages}{1841} (\bibinfo{year}{2017}).
\newblock \urlprefix\url{https://www.nature.com/articles/s41467-017-01882-w}.

\bibitem{ludescher_very_2014}
\bibinfo{author}{Ludescher, J.} \emph{et~al.}
\newblock \bibinfo{title}{Very early warning of next el nino}.
\newblock \emph{\bibinfo{journal}{Proceedings of the National Academy of
  Sciences}} \textbf{\bibinfo{volume}{111}}, \bibinfo{pages}{2064--2066}
  (\bibinfo{year}{2014}).
\newblock \urlprefix\url{http://www.pnas.org/content/111/6/2064}.

\bibitem{boers_prediction_2014}
\bibinfo{author}{Boers, N.} \emph{et~al.}
\newblock \bibinfo{title}{Prediction of extreme floods in the eastern {Central}
  {Andes} based on a complex networks approach}.
\newblock \emph{\bibinfo{journal}{Nature Communications}}
  \textbf{\bibinfo{volume}{5}}, \bibinfo{pages}{5199} (\bibinfo{year}{2014}).
\newblock \urlprefix\url{http://www.nature.com/doifinder/10.1038/ncomms6199}.

\bibitem{meng_percolation_2017}
\bibinfo{author}{Meng, J.}, \bibinfo{author}{Fan, J.},
  \bibinfo{author}{Ashkenazy, Y.} \& \bibinfo{author}{Havlin, S.}
\newblock \bibinfo{title}{Percolation framework to describe {El} {Nino}
  conditions}.
\newblock \emph{\bibinfo{journal}{Chaos: An Interdisciplinary Journal of
  Nonlinear Science}} \textbf{\bibinfo{volume}{27}}, \bibinfo{pages}{035807}
  (\bibinfo{year}{2017}).
\newblock \urlprefix\url{http://aip.scitation.org/doi/abs/10.1063/1.4975766}.

\bibitem{meng_forecasting_2018}
\bibinfo{author}{Meng, J.}, \bibinfo{author}{Fan, J.},
  \bibinfo{author}{Ashkenazy, Y.}, \bibinfo{author}{Bunde, A.} \&
  \bibinfo{author}{Havlin, S.}
\newblock \bibinfo{title}{Forecasting the magnitude and onset of {El} {Niño}
  based on climate network}.
\newblock \emph{\bibinfo{journal}{New Journal of Physics}}
  \textbf{\bibinfo{volume}{20}}, \bibinfo{pages}{043036}
  (\bibinfo{year}{2018}).
\newblock \urlprefix\url{http://stacks.iop.org/1367-2630/20/i=4/a=043036}.

\bibitem{newman2010networks}
\bibinfo{author}{Newman, M.}
\newblock \emph{\bibinfo{title}{Networks: an introduction}}
  (\bibinfo{publisher}{Oxford university press}, \bibinfo{year}{2010}).

\bibitem{cohen2010complex}
\bibinfo{author}{Cohen, R.} \& \bibinfo{author}{Havlin, S.}
\newblock \emph{\bibinfo{title}{Complex networks: structure, robustness and
  function}} (\bibinfo{publisher}{Cambridge university press},
  \bibinfo{year}{2010}).

\bibitem{zhao_inducing_2013}
\bibinfo{author}{Zhao, J.-H.}, \bibinfo{author}{Zhou, H.-J.} \&
  \bibinfo{author}{Liu, Y.-Y.}
\newblock \bibinfo{title}{Inducing effect on the percolation transition in
  complex networks}.
\newblock \emph{\bibinfo{journal}{Nature Communications}}
  \textbf{\bibinfo{volume}{4}}, \bibinfo{pages}{2412} (\bibinfo{year}{2013}).
\newblock
  \urlprefix\url{http://www.nature.com/ncomms/2013/130909/ncomms3412/full/ncomms3412.html}.

\bibitem{bardoscia_pathways_2017}
\bibinfo{author}{Bardoscia, M.}, \bibinfo{author}{Battiston, S.},
  \bibinfo{author}{Caccioli, F.} \& \bibinfo{author}{Caldarelli, G.}
\newblock \bibinfo{title}{Pathways towards instability in financial networks}.
\newblock \emph{\bibinfo{journal}{Nature Communications}}
  \textbf{\bibinfo{volume}{8}}, \bibinfo{pages}{14416} (\bibinfo{year}{2017}).
\newblock \urlprefix\url{https://www.nature.com/articles/ncomms14416}.

\bibitem{miura_effect_2012}
\bibinfo{author}{Miura, W.}, \bibinfo{author}{Takayasu, H.} \&
  \bibinfo{author}{Takayasu, M.}
\newblock \bibinfo{title}{Effect of {Coagulation} of {Nodes} in an {Evolving}
  {Complex} {Network}}.
\newblock \emph{\bibinfo{journal}{Physical Review Letters}}
  \textbf{\bibinfo{volume}{108}}, \bibinfo{pages}{168701}
  (\bibinfo{year}{2012}).
\newblock
  \urlprefix\url{https://link.aps.org/doi/10.1103/PhysRevLett.108.168701}.

\bibitem{acemoglu_systemic_2015}
\bibinfo{author}{Acemoglu, D.}, \bibinfo{author}{Ozdaglar, A.} \&
  \bibinfo{author}{Tahbaz-Salehi, A.}
\newblock \bibinfo{title}{Systemic {Risk} and {Stability} in {Financial}
  {Networks}}.
\newblock \emph{\bibinfo{journal}{American Economic Review}}
  \textbf{\bibinfo{volume}{105}}, \bibinfo{pages}{564--608}
  (\bibinfo{year}{2015}).
\newblock
  \urlprefix\url{https://www.aeaweb.org/articles?id=10.1257/aer.20130456}.

\bibitem{carmi_model_2007}
\bibinfo{author}{Carmi, S.}, \bibinfo{author}{Havlin, S.},
  \bibinfo{author}{Kirkpatrick, S.}, \bibinfo{author}{Shavitt, Y.} \&
  \bibinfo{author}{Shir, E.}
\newblock \bibinfo{title}{A model of {Internet} topology using k-shell
  decomposition}.
\newblock \emph{\bibinfo{journal}{Proceedings of the National Academy of
  Sciences}} \textbf{\bibinfo{volume}{104}}, \bibinfo{pages}{11150--11154}
  (\bibinfo{year}{2007}).
\newblock \urlprefix\url{http://www.pnas.org/content/104/27/11150}.

\bibitem{kitsak_identification_2010}
\bibinfo{author}{Kitsak, M.} \emph{et~al.}
\newblock \bibinfo{title}{Identification of influential spreaders in complex
  networks}.
\newblock \emph{\bibinfo{journal}{Nature Physics}}
  \textbf{\bibinfo{volume}{6}}, \bibinfo{pages}{888--893}
  (\bibinfo{year}{2010}).
\newblock
  \urlprefix\url{http://www.nature.com/nphys/journal/v6/n11/full/nphys1746.html}.

\bibitem{breiman_random_2001}
\bibinfo{author}{Breiman, L.}
\newblock \bibinfo{title}{Random {Forests}}.
\newblock \emph{\bibinfo{journal}{Machine Learning}}
  \textbf{\bibinfo{volume}{45}}, \bibinfo{pages}{5--32} (\bibinfo{year}{2001}).
\newblock \urlprefix\url{https://doi.org/10.1023/A:1010933404324}.

\bibitem{hoberg_product_2010}
\bibinfo{author}{Hoberg, G.} \& \bibinfo{author}{Phillips, G.}
\newblock \bibinfo{title}{Product {Market} {Synergies} and {Competition} in
  {Mergers} and {Acquisitions}: {A} {Text}-{Based} {Analysis}}.
\newblock \emph{\bibinfo{journal}{Review of Financial Studies}}
  \textbf{\bibinfo{volume}{23}}, \bibinfo{pages}{3773--3811}
  (\bibinfo{year}{2010}).
\newblock \urlprefix\url{http://rfs.oxfordjournals.org/content/23/10/3773}.

\bibitem{Hoberg2016jpe}
\bibinfo{author}{Hoberg, G.} \& \bibinfo{author}{Phillips, G.}
\newblock \bibinfo{title}{Text-based network industries and endogenous product
  differentiation}.
\newblock \emph{\bibinfo{journal}{Journal of Political Economy}}
  \textbf{\bibinfo{volume}{124}}, \bibinfo{pages}{1423--1465}
  (\bibinfo{year}{2016}).
\newblock \urlprefix\url{https://doi.org/10.1086/688176}.
\newblock \eprint{https://doi.org/10.1086/688176}.

\bibitem{fama1992cross}
\bibinfo{author}{Fama, E.~F.} \& \bibinfo{author}{French, K.~R.}
\newblock \bibinfo{title}{The cross-section of expected stock returns}.
\newblock \emph{\bibinfo{journal}{the Journal of Finance}}
  \textbf{\bibinfo{volume}{47}}, \bibinfo{pages}{427--465}
  (\bibinfo{year}{1992}).

\bibitem{cooper2008asset}
\bibinfo{author}{Cooper, M.~J.}, \bibinfo{author}{Gulen, H.} \&
  \bibinfo{author}{Schill, M.~J.}
\newblock \bibinfo{title}{Asset growth and the cross-section of stock returns}.
\newblock \emph{\bibinfo{journal}{The Journal of Finance}}
  \textbf{\bibinfo{volume}{63}}, \bibinfo{pages}{1609--1651}
  (\bibinfo{year}{2008}).

\bibitem{amihud2002illiquidity}
\bibinfo{author}{Amihud, Y.}
\newblock \bibinfo{title}{Illiquidity and stock returns: cross-section and
  time-series effects}.
\newblock \emph{\bibinfo{journal}{Journal of financial markets}}
  \textbf{\bibinfo{volume}{5}}, \bibinfo{pages}{31--56} (\bibinfo{year}{2002}).

\bibitem{altman1968financial}
\bibinfo{author}{Altman, E.~I.}
\newblock \bibinfo{title}{Financial ratios, discriminant analysis and the
  prediction of corporate bankruptcy}.
\newblock \emph{\bibinfo{journal}{The journal of finance}}
  \textbf{\bibinfo{volume}{23}}, \bibinfo{pages}{589--609}
  (\bibinfo{year}{1968}).

\bibitem{rosen_topological_2016}
\bibinfo{author}{Rosen, Y.} \& \bibinfo{author}{Louzoun, Y.}
\newblock \bibinfo{title}{Topological similarity as a proxy to content
  similarity}.
\newblock \emph{\bibinfo{journal}{Journal of Complex Networks}}
  \textbf{\bibinfo{volume}{4}}, \bibinfo{pages}{38--60} (\bibinfo{year}{2016}).
\newblock
  \urlprefix\url{https://academic.oup.com/comnet/article/4/1/38/2366087}.

\bibitem{naaman_edge_nodate}
\bibinfo{author}{Naaman, R.}, \bibinfo{author}{Cohen, K.} \&
  \bibinfo{author}{Louzoun, Y.}
\newblock \bibinfo{title}{Edge sign prediction based on a combination of
  network structural topology and sign propagation}.
\newblock \emph{\bibinfo{journal}{Journal of Complex Networks}}
  \urlprefix\url{https://academic.oup.com/comnet/advance-article/doi/10.1093/comnet/cny012/4999727}.

\bibitem{everett_centrality_1999}
\bibinfo{author}{Everett, M.~G.} \& \bibinfo{author}{Borgatti, S.~P.}
\newblock \bibinfo{title}{The centrality of groups and classes}.
\newblock \emph{\bibinfo{journal}{The Journal of Mathematical Sociology}}
  \textbf{\bibinfo{volume}{23}}, \bibinfo{pages}{181--201}
  (\bibinfo{year}{1999}).
\newblock \urlprefix\url{https://doi.org/10.1080/0022250X.1999.9990219}.

\bibitem{sabidussi_centrality_1966}
\bibinfo{author}{Sabidussi, G.}
\newblock \bibinfo{title}{The centrality index of a graph}.
\newblock \emph{\bibinfo{journal}{Psychometrika}}
  \textbf{\bibinfo{volume}{31}}, \bibinfo{pages}{581--603}
  (\bibinfo{year}{1966}).
\newblock \urlprefix\url{https://doi.org/10.1007/BF02289527}.

\bibitem{dijkstra_note_1959}
\bibinfo{author}{Dijkstra, E.~W.}
\newblock \bibinfo{title}{A {Note} on {Two} {Problems} in {Connexion} with
  {Graphs}}.
\newblock \emph{\bibinfo{journal}{Numer. Math.}} \textbf{\bibinfo{volume}{1}},
  \bibinfo{pages}{269--271} (\bibinfo{year}{1959}).
\newblock \urlprefix\url{http://dx.doi.org/10.1007/BF01386390}.

\bibitem{rosen_directionality_2014}
\bibinfo{author}{Rosen, Y.} \& \bibinfo{author}{Louzoun, Y.}
\newblock \bibinfo{title}{Directionality of real world networks as predicted by
  path length in directed and undirected graphs}.
\newblock \emph{\bibinfo{journal}{Physica A: Statistical Mechanics and its
  Applications}} \textbf{\bibinfo{volume}{401}}, \bibinfo{pages}{118--129}
  (\bibinfo{year}{2014}).
\newblock
  \urlprefix\url{http://www.sciencedirect.com/science/article/pii/S0378437114000090}.

\bibitem{milo_network_2002}
\bibinfo{author}{Milo, R.} \emph{et~al.}
\newblock \bibinfo{title}{Network {Motifs}: {Simple} {Building} {Blocks} of
  {Complex} {Networks}}.
\newblock \emph{\bibinfo{journal}{Science}} \textbf{\bibinfo{volume}{298}},
  \bibinfo{pages}{824--827} (\bibinfo{year}{2002}).
\newblock \urlprefix\url{http://science.sciencemag.org/content/298/5594/824}.

\bibitem{itzhack_optimal_2007}
\bibinfo{author}{Itzhack, R.}, \bibinfo{author}{Mogilevski, Y.} \&
  \bibinfo{author}{Louzoun, Y.}
\newblock \bibinfo{title}{An optimal algorithm for counting network motifs}.
\newblock \emph{\bibinfo{journal}{Physica A: Statistical Mechanics and its
  Applications}} \textbf{\bibinfo{volume}{381}}, \bibinfo{pages}{482--490}
  (\bibinfo{year}{2007}).
\newblock
  \urlprefix\url{http://www.sciencedirect.com/science/article/pii/S0378437107002257}.

\bibitem{batagelj_fast_2011}
\bibinfo{author}{Batagelj, V.} \& \bibinfo{author}{Zaveršnik, M.}
\newblock \bibinfo{title}{Fast algorithms for determining (generalized) core
  groups in social networks}.
\newblock \emph{\bibinfo{journal}{Advances in Data Analysis and
  Classification}} \textbf{\bibinfo{volume}{5}}, \bibinfo{pages}{129--145}
  (\bibinfo{year}{2011}).
\newblock \urlprefix\url{https://doi.org/10.1007/s11634-010-0079-y}.

\bibitem{ho_random_1995}
\bibinfo{author}{Ho, T.~K.}
\newblock \bibinfo{title}{Random decision forests}.
\newblock In \emph{\bibinfo{booktitle}{Proceedings of 3rd {International}
  {Conference} on {Document} {Analysis} and {Recognition}}},
  vol.~\bibinfo{volume}{1}, \bibinfo{pages}{278--282 vol.1}
  (\bibinfo{year}{1995}).

\bibitem{kingma_adam:_2014}
\bibinfo{author}{Kingma, D.~P.} \& \bibinfo{author}{Ba, J.}
\newblock \bibinfo{title}{Adam: {A} {Method} for {Stochastic} {Optimization}}.
\newblock \emph{\bibinfo{journal}{arXiv:1412.6980 [cs]}}
  (\bibinfo{year}{2014}).
\newblock \urlprefix\url{http://arxiv.org/abs/1412.6980}.
\newblock \bibinfo{note}{ArXiv: 1412.6980}.

\bibitem{SI}
\emph{\bibinfo{journal}{Supplementary materials}} .

\bibitem{watts_collective_1998}
\bibinfo{author}{Watts, D.~J.} \& \bibinfo{author}{Strogatz, S.~H.}
\newblock \bibinfo{title}{Collective dynamics of ‘small-world’ networks}.
\newblock \emph{\bibinfo{journal}{Nature}} \textbf{\bibinfo{volume}{393}},
  \bibinfo{pages}{440--442} (\bibinfo{year}{1998}).
\newblock
  \urlprefix\url{http://www.nature.com/nature/journal/v393/n6684/full/393440a0.html}.

\bibitem{barrat_architecture_2004}
\bibinfo{author}{Barrat, A.}, \bibinfo{author}{Barthélemy, M.},
  \bibinfo{author}{Pastor-Satorras, R.} \& \bibinfo{author}{Vespignani, A.}
\newblock \bibinfo{title}{The architecture of complex weighted networks}.
\newblock \emph{\bibinfo{journal}{Proceedings of the National Academy of
  Sciences of the United States of America}} \textbf{\bibinfo{volume}{101}},
  \bibinfo{pages}{3747--3752} (\bibinfo{year}{2004}).
\newblock \urlprefix\url{http://www.pnas.org/content/101/11/3747}.

\bibitem{saramaki_generalizations_2007}
\bibinfo{author}{Saramäki, J.}, \bibinfo{author}{Kivelä, M.},
  \bibinfo{author}{Onnela, J.-P.}, \bibinfo{author}{Kaski, K.} \&
  \bibinfo{author}{Kertész, J.}
\newblock \bibinfo{title}{Generalizations of the clustering coefficient to
  weighted complex networks}.
\newblock \emph{\bibinfo{journal}{Physical Review E}}
  \textbf{\bibinfo{volume}{75}}, \bibinfo{pages}{027105}
  (\bibinfo{year}{2007}).
\newblock \urlprefix\url{https://link.aps.org/doi/10.1103/PhysRevE.75.027105}.

\end{thebibliography}
\appendix

%%%%%%%%%% Merge with supplemental materials %%%%%%%%%%
\pagebreak
\widetext
\begin{center}
\textbf{\large Supplemental Materials: A network and machine learning approach to product market forecasting}

Jingfang Fan, Keren  Cohen, Louis M. Shekhtman,  Sibo Liu,
Jun Meng,  Yoram Louzoun,  and Shlomo Havlin
\end{center}

\setcounter{equation}{0}
\setcounter{figure}{0}
\setcounter{table}{0}
%\makeatletter
\renewcommand{\theequation}{S\arabic{equation}}
\renewcommand{\thefigure}{S\arabic{figure}}

%\begin{equation}
%\left\{
%\begin{array}{ll}
%\boldsymbol{x}_{<}  =  \boldsymbol{z}_{<},  \\ 
%\boldsymbol{x}_{>}   =  \boldsymbol{z}_{>} \odot e^{s(\boldsymbol{z}_{<})} + t(\boldsymbol{z}_{<}), \label{eq:rnvp}
%\end{array}
%\right.
%\end{equation}
%The transformation \Eq{eq:rnvp} is easy to invert by reversing the basic arithmetical

\section{Further results}
We present here further results not shown in the main text.

We first prove that the PDF of $w_{i,j}$ [see, Eq. \ref{eq4}] follows an exponential distribution by using a convolution method.
We assume $f_{i}$ to be the full density function of $x_{i}$, and let $f_{j}$ be the full density function of $x_{j}$. Here, we assume that $f_{i}$ and $f_{j}$ are uniform random distributions.
Then, for case 1, $x_{j} \geq x_{i}$,
\begin{eqnarray}
f_{w}(w_{i,j}) &=& \int_{-\infty}^{\infty} f_{i}(x_{i}) f_{j}(x_{i} +1 - e^{-\frac{w{i,j}}{\Lambda}}) d x_{i}  \nonumber \\
&=&\int_{0}^{1} f_{j}(x_{i} +1 - e^{-\frac{w{i,j}}{\Lambda}}) d x_{i}.  \nonumber \\
\label{Seq1}
\end{eqnarray}
This integral is zero unless  $0 \leq x_{i} +1 - e^{-\frac{w{i,j}}{\Lambda}} \leq 1$, implying $-1 + e^{-\frac{w{i,j}}{\Lambda}} \leq x_{i} \leq  e^{-\frac{w{i,j}}{\Lambda}}$,
since $0 \leq e^{-\frac{w{i,j}}{\Lambda}} \leq 1$, therefore,

\begin{eqnarray}
f_{w}(w_{i,j}) &=&  \int_{0}^{e^{-\frac{w{i,j}}{\Lambda}} } d x_{i}  \nonumber \\
&=& e^{-\frac{w{i,j}}{\Lambda}}.  \nonumber \\
\label{eq3}
\end{eqnarray}

For case 2, $x_{i} > x_{j}$,

\begin{eqnarray}
f_{w}(w_{i,j}) &=& \int_{-\infty}^{\infty} f_{i}(x_{i}) f_{j}(x_{i} -1 + e^{-\frac{w{i,j}}{\Lambda}}) d x_{i}  \nonumber \\
&=&\int_{0}^{1} f_{j}(x_{i} -1 + e^{-\frac{w{i,j}}{\Lambda}}) d x_{i},  \nonumber \\
\label{eq4}
\end{eqnarray}
this integral is zero unless  $0 \leq x_{i} -1 + e^{-\frac{w{i,j}}{\Lambda}} \leq 1$, implying $1 - e^{-\frac{w{i,j}}{\Lambda}} \leq x_{i} \leq 2 - e^{-\frac{w{i,j}}{\Lambda}}$,
since $0 \leq e^{-\frac{w{i,j}}{\Lambda}} \leq 1$. Thus,

\begin{eqnarray}
f_{w}(w_{i,j}) &=&  \int_{1 - e^{-\frac{w{i,j}}{\Lambda}}}^{1} d x_{i}  \nonumber \\
&=& e^{-\frac{w{i,j}}{\Lambda}}.  \nonumber \\
\label{eq5}
\end{eqnarray}

From Eq. \ref{eq3} and \ref{eq5} we obtain that the probability distribution functions of link weights  in our model $p(w)$ follows an exponential distribution.

\clearpage
\begin{figure}
\begin{centering}
\includegraphics[width=1.0\linewidth]{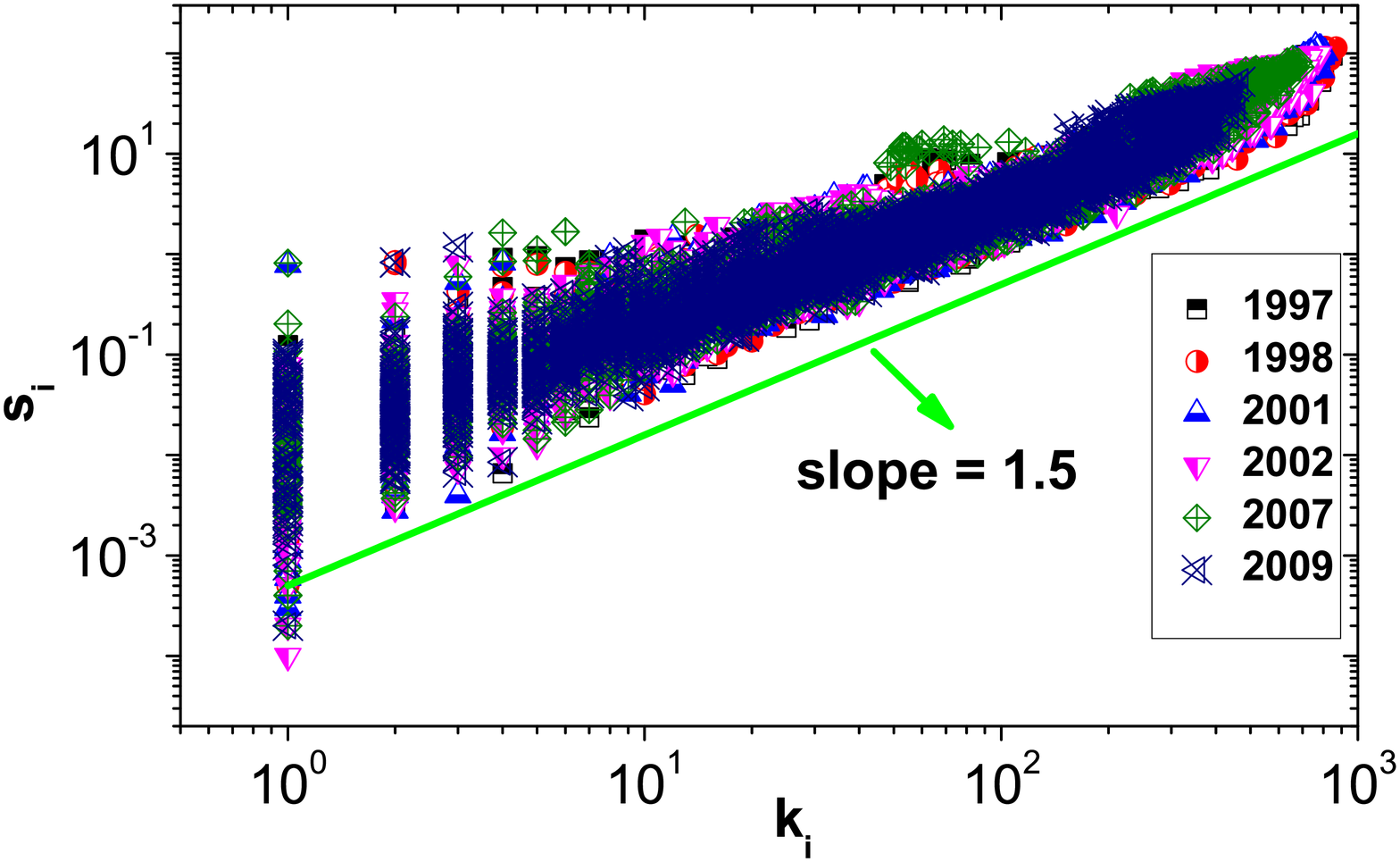}
\caption{\label{Fig:S1} (Strength $s$ as function of the degree $k$ of nodes.  The line shows that the real data follow a power law behavior with an exponent $ \beta \approx 1.5$.  }
\end{centering}
\end{figure}

\begin{figure}
\begin{centering}
\includegraphics[width=1\textwidth]{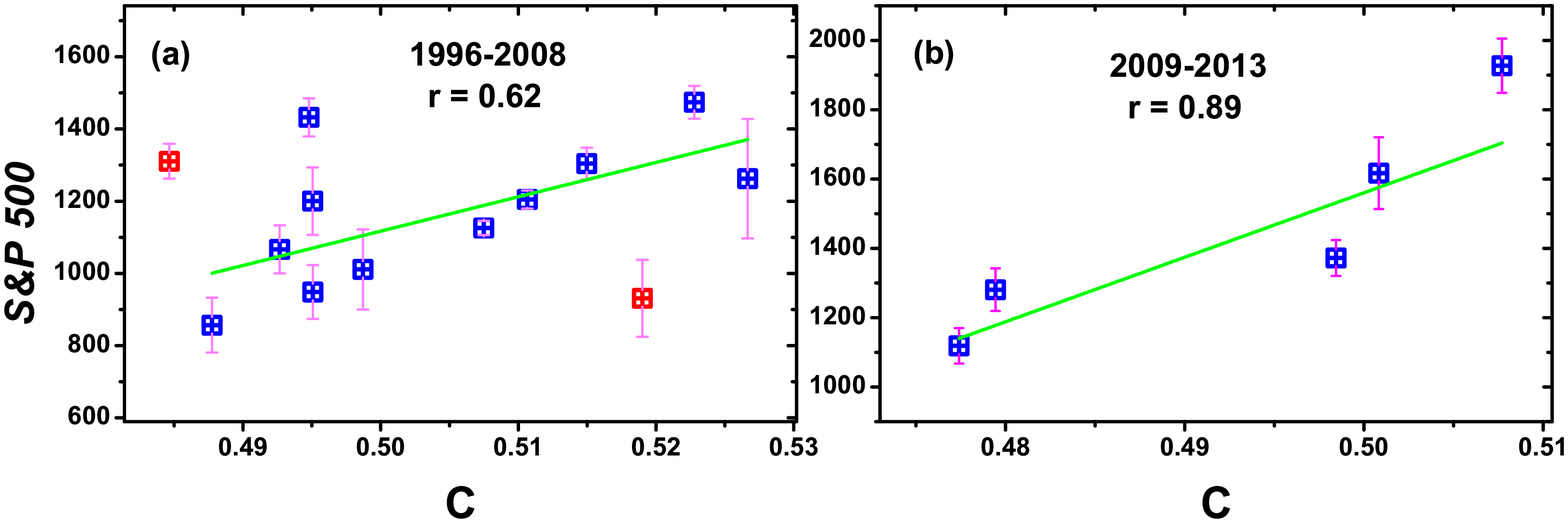}
\caption{\label{Figs:S2} The network Index $C$  versus the annual average S\&P 500 Index of next year. (a) The years from 1996 to 2008, i.e before financial crisis. (b) The years from 2009 to 2013, i.e after financial crisis. The straight lines show the best linear regression with correlation $r$.}
\par\end{centering}
\end{figure}

\begin{figure}
\begin{centering}
\includegraphics[width=1\textwidth]{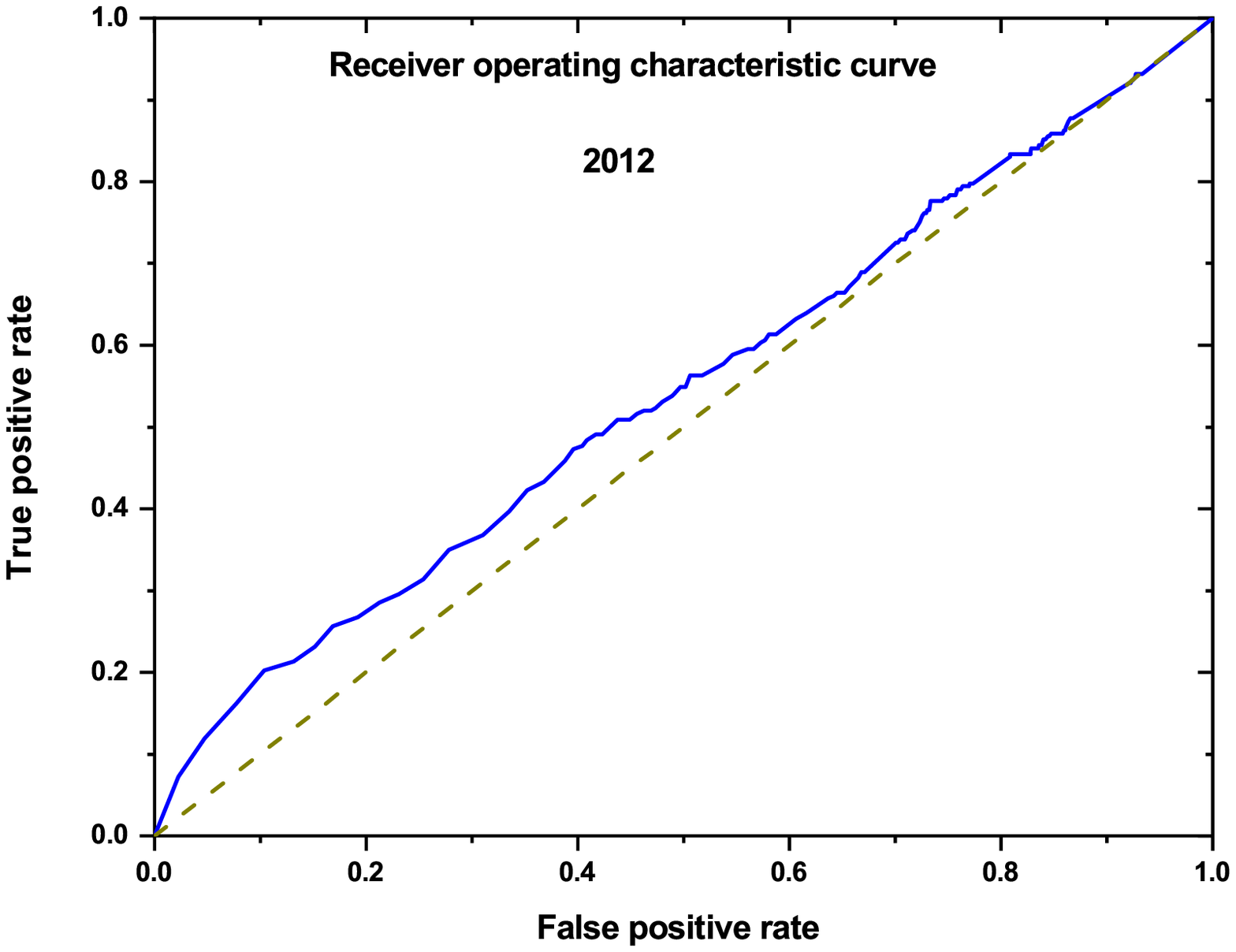}
\caption{\label{Figs:S3}  The receiver operating characteristic curve (ROC) using the k-shell method for predicting the disappeared firms in the 2012 network.}
\par\end{centering}
\end{figure}

\begin{figure}
\begin{centering}
\includegraphics[width=0.7\textwidth]{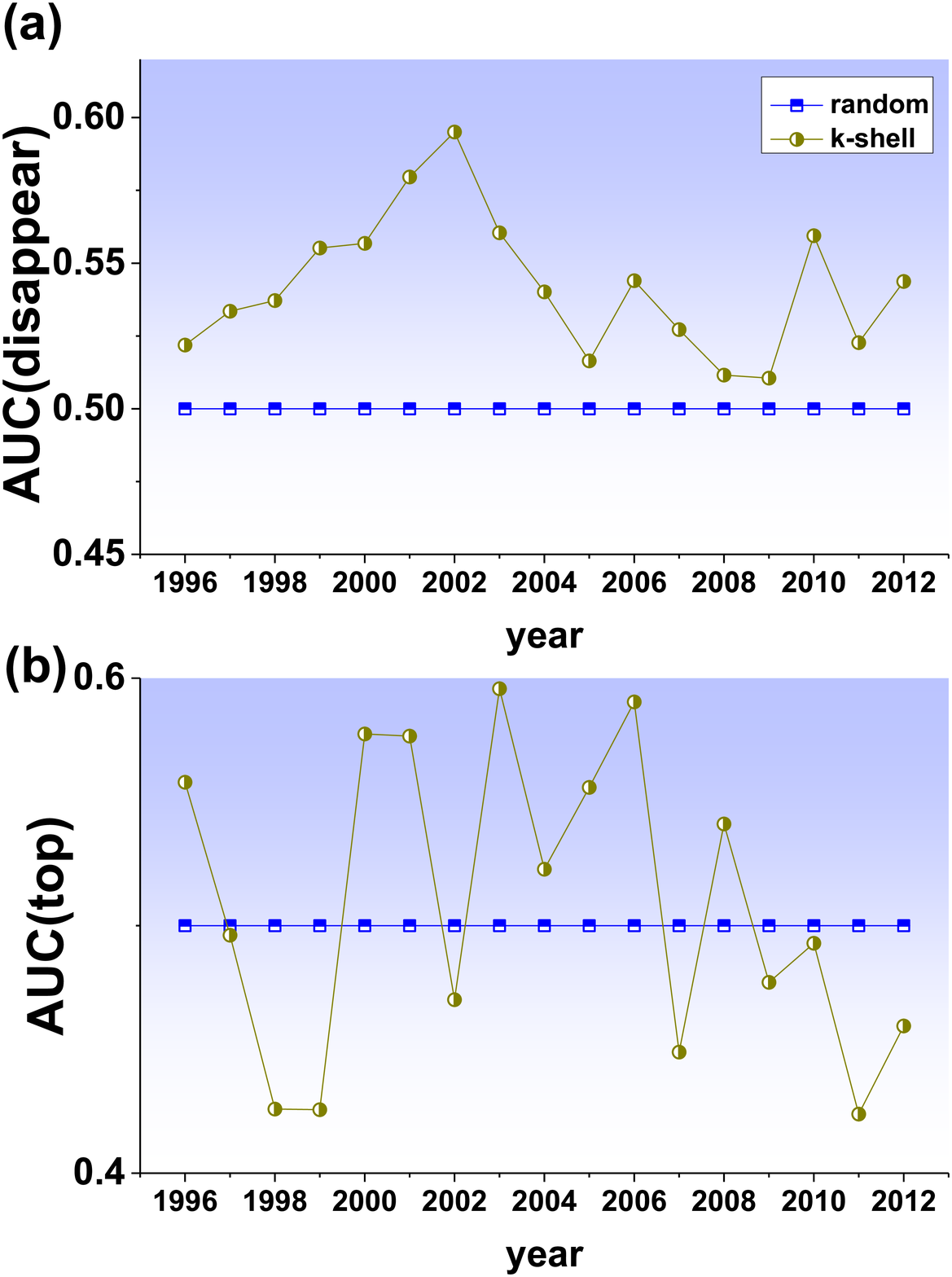}
\caption{\label{Figs:S4} The area under the ROC curve for (a) disappeared firms and (b) top return firms using forecasts from the $K$-shell methods.}
\par\end{centering}
\end{figure}

\begin{figure}
\begin{centering}
\includegraphics[width=0.7\textwidth]{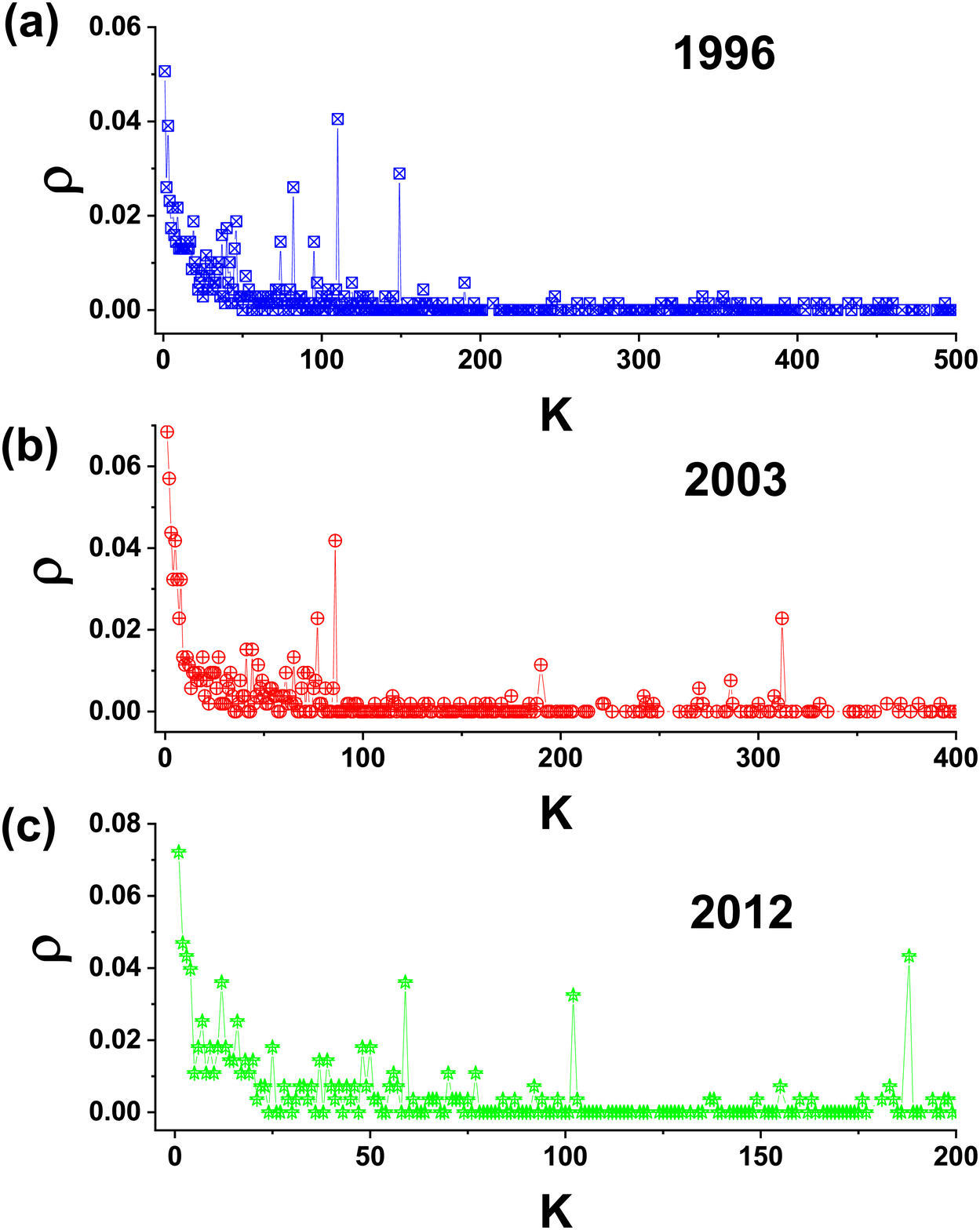}
\caption{\label{Figs:S5} The risk ratio, $\rho$, as a function of $K$-shell for (a) 1996, (b) 2003 and (c) 2012 years.}
\par\end{centering}
\end{figure}

\end{document}